\documentclass[multi]{cambridge6A}
\usepackage[numbers]{natbib}
\usepackage{rotating}
\usepackage{floatpag}
\rotfloatpagestyle{empty}
\usepackage{amsthm}
\usepackage{graphicx}

\usepackage{ubec}

\usepackage{amsmath}
\newcommand{\kk}{\mathbf{k}}
\newcommand{\rr}{\mathbf{r}}

\newcommand{\dd}{\mathrm{d}}
\newcommand{\ee}{\mathrm{e}}

\newcommand{\bh}{\hat{b}}
\newcommand{\bhd}{\hat{b}^\dagger}
\newcommand{\ah}{\hat{a}}
\newcommand{\ahd}{\hat{a}^\dagger}
\newcommand{\ch}{\hat{c}}
\newcommand{\chd}{\hat{c}^\dagger}

\newcommand{\Eh}{\hat{E}}
\newcommand{\Ehd}{\hat{E}^\dagger}
\newcommand{\Psih}{\hat{\Psi}}
\newcommand{\Psihd}{\hat{\Psi}^\dagger}
\newcommand{\eq}[1]{(\ref{eq:#1})}
\newcommand{\eqname}[1]{\label{eq:#1}}

\begin{document}

\author{Alessio Chiocchetta and Andrea Gambassi}
\affiliation{SISSA - International School for Advanced Studies and INFN, via Bonomea 265, 34136 Trieste, Italy}

\author{Iacopo Carusotto}
\affiliation{INO-CNR BEC Center and Dipartimento di Fisica, Universit\`a di Trento, Via Sommarive 14, I-38123 Povo, Italy}

\chapter{Laser operation and Bose-Einstein condensation: analogies and differences} 

\begin{abstract}
After reviewing the interpretation of laser operation as a non-equi\-li\-brium Bose-Einstein condensation phase transition, we illustrate the novel features arising from the non-equilibrium nature of photon and polariton Bose-Einstein condensates recently observed in experiments. We then propose a quantitative criterion to experimentally assess the equilibrium vs. non-equi\-li\-brium nature of a specific condensation process, based on fluctuation-dissipation relations. The power of this criterion is illustrated on two models which shows very different behaviours.
%
%
\end{abstract}

\section{Historical and conceptual introduction}
\label{carusotto_sec1}
 
The first introduction of non-equilibrium statistical mechanics concepts into the realm of optics dates back to the early 1970's with pioneering works by Graham and Haken~\cite{Graham:ZPhys1970} and by DeGiorgio and Scully~\cite{DeGiorgio:PRA1970}, who proposed a very insightful interpretation of the laser threshold in terms of a spontaneous breaking of the $U(1)$ symmetry associated with the phase of the emitted light. Similarly to what happens to the order parameter at a second-order phase transition, such an optical phase is randomly chosen every time the device is switched on and remains constant for macroscopic times. Moreover, a long-range spatial order is established, as light emitted by a laser device above threshold is phase-coherent on macroscopic distances.


While textbooks typically discuss this interpretation of laser operation in terms of a phase transition for the simplest case of a single-mode laser cavity, rigorously speaking this is valid only in spatially infinite systems.
%
%
In fact, only in this case one can observe non-analytic behaviors of the physical quantities at the transition point. In particular, the long-range order is typically assessed by looking at the long-distance behaviour of the correlation function of the order parameter, which, for a laser, corresponds to the first-order spatial coherence of the emitted electric field $\Eh(\rr)$,
\begin{equation}
\lim_{|\rr-\rr'|\to\infty} \left\langle \Ehd(\rr)\,\Eh(\rr') \right\rangle:
\end{equation}
the spontaneous symmetry breaking is signalled by this quantity becoming non-zero. The average $\langle \dots \rangle$ is taken on the stationary density matrix of the system. In order to be able to probe long-distance behaviour, experimental studies need devices with a spatially extended active region. The so-called VCSELs (vertical cavity surface emitting lasers) are perhaps the most studied examples in this class~\cite{VCSEL,Kapon}: by using an active medium sandwiched between a pair of plane-parallel semiconductor mirrors, one can realize devices of arbitrary size, the only limitation coming from extrinsic effects such as the difficulty of having a spatially homogeneous pumping of the active material and of 
avoiding disorder of the semiconductor microstructure.

The most celebrated phase transition breaking a $U(1)$ symmetry in statistical physics is perhaps the Bose-Einstein condensation (BEC). While in textbooks~\cite{Huang,BECbook} BEC is typically described in terms of the emergence of a macroscopic occupation of a single quantum level, which gives a finite condensate density $n_{BEC}$, an alternative, mathematically equivalent condition ---the so-called Penrose-Onsager criterion~\cite{penrose_onsager_56}--- involves the long-distance limit of the coherence function of the matter field $\Psih(\rr)$ describing the Bose particles undergoing condensation, i.e.,
\begin{equation}
 \lim_{|\rr-\rr'|\to\infty} \left\langle \Psihd(\rr)\,\Psih(\rr') \right\rangle = n_{BEC} > 0,
 \eqname{PO-BEC}
\end{equation}
where the average $\langle \dots\rangle $ is taken on the thermal density matrix. On this basis, it is natural to see BEC as a phase transition which spontaneously breaks the global $U(1)$ gauge symmetry of the quantum matter field $\Psih(\rr)\to \ee^{i \theta}\Psih(\rr)$.

There is however an important difference between textbook BEC and laser operation: in the former, the system is assumed to be in thermal equilibrium at temperature $T$, so the density matrix $\rho_{\rm eq}$ of the system is given by the Boltzmann factor of equilibrium statistical mechanics $\rho_{\rm eq} \propto \exp(-H/k_B T )$. A laser is, instead, an intrinsically non-equilibrium device, whose steady state is determined by a dynamical balance of pumping and losses, the latter being essential to generate the output laser beam used in any application. As a result, one can think of laser operation in spatially extended devices as an example of {\em non-equilibrium BEC}. In between the two extreme limits of equilibrium BEC and laser operation, experiments with gases of exciton-polaritons in microcavities~\cite{Kasprzak}, of magnons~\cite{Demokritov:Nature2006} and of photons~\cite{klaers_schmitt_10} have explored a full range of partially thermalized regimes depending on the ratio between the loss and the thermalization rates, the latter being typically due to interparticle collisions within the gas and/or to interactions with the host material.

The first part of this Chapter will be devoted to a brief review of the novel features of non-equilibrium BEC as compared to its equilibrium counterpart. In the second part we will dwell on the equilibrium vs. non-equilibrium character of this phenomena by proposing a quantitative criterion to experimentally probe the nature of the stationary state in specific cases, based on the fluctuation-dissipation theorems. In order to illustrate the practical utility of this approach, two toy models of condensation will be discussed.


\section{Textbook equilibrium  Bose-Einstein condensation}
\label{carusotto_sec2}

The textbook discussion of the equilibrium BEC in statistical physics~\cite{Huang,BECbook} is based on a grand-canonical description of an ideal three-dimensional Bose gas at thermal equilibrium with inverse temperature $\beta=1/k_B T$ and chemical potential $\mu \leq 0$ in terms of the Bose distribution, $n_\kk=[e^{\beta(\epsilon_\kk-\mu)}-1]^{-1}$, where $\epsilon_\kk=\hbar^2 k^2 /2m$ is the energy of the state of momentum $\hbar\kk$, and $m$ the mass of the particles.

For any given temperature $T$, the maximum density of particles that can be accommodated in the excited states at $\kk\neq 0$ grows with $\mu$ and then saturates at $n_{\rm max}(T)$ for $\mu=0$.
%
%
If the actual density of particles $n$ exceeds this threshold, the extra particles must accumulate into the lowest state at $\kk=0$, forming the so-called condensate. 
%
%
%
As the coherence in Eq.~\eq{PO-BEC} is the Fourier transform of the momentum distribution $n_\kk$, it is immediate to see that the two different definitions of condensate fraction $n_{BEC}$ in terms of a macroscopic occupation $n_{\kk=0}$ of the lowest mode and of the long-distance behaviour of the coherence actually coincide.

An alternative but equivalent description of this physics was presented in Ref.~\cite{Gunton:PR1968}: in order to highlight the spontaneous symmetry breaking mechanism it is convenient to introduce a fictitious external field $\eta $ coupling to the quantum matter field via the Hamiltonian
\begin{equation}
 H_{\rm ext}=-\int\,d^3\rr\,\left[\eta \Psihd(\rr) + \eta^* \Psih(\rr)\right],
\end{equation}
which explicitly breaks the $U(1)$ symmetry. 
As in a ferromagnet the direction of the magnetization is selected by an external magnetic field $B$,
%
%
in the presence of an $\eta$ field the quantum matter field $\Psih(\rr)$ acquires a finite expectation value. For $T<T_{BEC}$, this expectation value 
\begin{equation}
\label{eq:order-parameter}
\Psi_0= \lim_{\eta\to 0} \lim_{V\to \infty} \left \langle \Psih(\rr) \right\rangle
\end{equation}
remains finite even for vanishing $\eta$ in the thermodynamic limit $V\to\infty$, and is related to the condensate density by $n_{BEC}=|\Psi_0|^2$. As usual for phase transitions, the order in which the zero-field and the infinite-volume limits are taken in Eq.~\eqref{eq:order-parameter} is crucial~\cite{Huang}.

In the presence of inter-particle pair interactions, all condensation criteria based on the lowest mode occupation, on the long-distance coherence and on the spontaneous coherent matter field remain valid. However, the underlying physics is much more complex and we refer the reader to the specialized literature on the subject, e.g. Ref.~\cite{BECbook}. For our purposes, we only need to mention that at equilibrium at $T=0$ the condensate order parameter $\Psi_0(\rr)$ in presence of an external potential $V(\rr)$ can be obtained in the dilute gas regime by minimizing the so-called Gross-Pitaevskii (GP) energy functional,
\begin{equation}
 E[\phi_0]=\int\!d^3\rr\, \left\{\frac{\hbar^2 |\nabla\Psi_0(\rr)|^2}{2m} + V(\rr) |\Psi_0(\rr)|^2 + \frac{g}{2} |\Psi_0(\rr)|^4\right\},
\eqname{carusotto_GP-static}
 \end{equation}
where the normalization of $\Psi_0$ is set to the total number $N$ of particles in the system, 
$N=\int\!d^3\rr\,|\Psi_0(\rr)|^2$
and $g=4\pi \hbar^2 a_0/m$ quantifies the strength of the (local) interactions proportional to the $s$-wave inter-particle scattering length $a_0$. In terms of $a_0$, the dilute gas regime corresponds to $n a_0^3 \ll 1$. 
The mathematical form of the GP energy functional \eq{carusotto_GP-static} guarantees that the condensate wavefunction $\Psi_0(\rr)$ keeps a constant phase throughout the whole system  even in the presence of the external potential $V(\rr)$, while the density profile $|\Psi_0(\rr)|^2$ develops large variations in space.

Finally, at this same level of approximation, the condensate dynamics is ruled by the time-dependent Gross-Pitaevskii equation,
\begin{equation}
i\hbar \frac{\partial \Psi_0(\rr,t)}{\partial t}=-\frac{\hbar^2 \nabla^2 \Psi_0(\rr,t)}{2m} + V(\rr) \Psi_0(\rr,t) + g |\Psi_0(\rr,t)|^2\,\Psi_0(\rr,t) ,
\eqname{carusotto_GP-dynamic}
\end{equation}
which has the mathematical form of a nonlinear Schr\"odinger equation.

\section{Mean-field theory of non-equilibrium BEC}

While the shape of an equilibrium condensate is obtained by minimizing the energy functional \eq{carusotto_GP-static} (of the Ginzburg-Landau type), non-equilibrium BECs appear to be significantly less universal as their theoretical description typically requires some microscopic modelling of the specific dissipation and pumping mechanisms present in a given experimental set-up. For a comprehensive discussion of the main configurations, we refer the interested reader to the recent review article~\cite{RMP}. 
Here we summarize the simplest and most transparent of such descriptions, which is based on the following complex Ginzburg-Landau evolution equation for the order parameter,
\begin{equation}
 i\hbar \frac{\partial \Psi_0}{\partial t}=-\frac{\hbar^2  }{2m}\nabla^2\Psi_0 + V(\rr) \Psi_0 + g |\Psi_0|^2\,\Psi_0-\frac{i\hbar}{2}\left[ \gamma - \frac{P}{1+\frac{|\Psi_0|^2}{n_{\rm sat}}}\right]\Psi_0,
\eqname{carusotto_CGLE}
\end{equation}
inspired from the so-called semiclassical theory of the laser. In addition to the terms already present in the equilibrium description \eq{carusotto_GP-dynamic}, Eq. \eq{carusotto_CGLE} accounts for the losses at a rate $\gamma$ and for the stimulated pumping of new particles into the condensate at a bare rate $P$, which then saturates once the condensate density exceeds the saturation density $n_{\rm sat}$. Given the driven-dissipative nature of this evolution equation, the steady state has to be determined as the long-time limit of the dynamical evolution.

This apparently minor difference has profound implications, as the breaking of time-reversal symmetry by the pumping and loss terms in \eq{carusotto_CGLE} allows for steady-state configurations with a spatially varying phase of the order parameter, which physically corresponds to finite particle currents through the condensate. This feature was first observed in Ref.~\cite{Richard_small} as a ring-shaped condensate emission in the wavevector $\kk$-space and, in the presence of disorder, as an asymmetry of the emission pattern under reflections, $n(\kk)\neq n(-\kk)$~\cite{Richard_big}. A theoretical interpretation was proposed in Ref.~\cite{WCC} and soon confirmed by the more detailed experiments in Ref.~\cite{Wertz}.

Another feature of Eq. \eq{carusotto_CGLE} which clearly distinguishes non-equilibrium systems from their equilibrium counterparts is the dispersion of the collective excitations on top of a condensate. As first predicted in Refs.~\cite{Wouters:PRB2006,Wouters:PRA2007,Szymanska:PRL2006}, the usual Bogoliubov dispersion of spatially homogeneous condensates
$\hbar \omega_{{\rm eq},\kk} = [\epsilon_\kk(\epsilon_\kk+2 g|\Psi_0|^2)]^{1/2}$
is modified by pumping and dissipation to
$\omega_{{\rm neq},\kk} =-{i\Gamma}/{2}+[{\omega_{{\rm eq},\kk}^2 -{\Gamma^2}/{4}]^{1/2}},$
where the dissipation parameter $\Gamma = \gamma (1- P_c/P)$ depends on the pumping power $P$ in units of its threshold value $P_c$. 
In particular, the usual sonic (linear) dispersion of low-wavevector excitations in equilibrium condensates is strongly modified into a flat, diffusive region~\cite{Wouters:PRL2007}.


\section{Non-condensed cloud and quasi-condensation}

\begin{figure}[htbp]
\centerline{\begin{tabular}{cc}
	\includegraphics[width=6.cm]{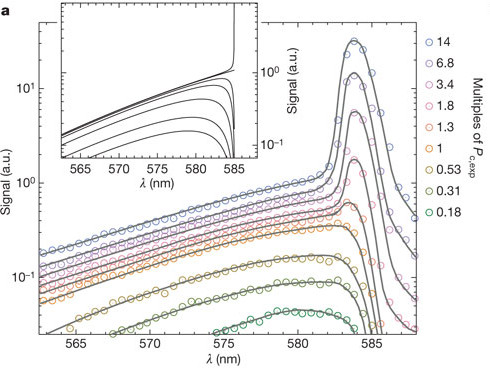} &
	\includegraphics[width=3.2cm]{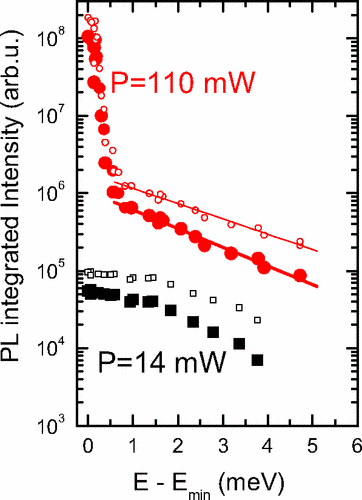}
\end{tabular}}
\caption{Left panel: wavelength distribution of the emission from a photon condensate across the condensation threshold ( from Ref.~\cite{klaers_schmitt_10}).
Right panel: energy distribution of the emission from a microcavity for pump values $P$ respectively below (black) and above (red) the condensation threshold. In the latter case, photons are expected to be almost non-interacting particles (from Ref.~\cite{Bajoni}). }
\label{carusotto_fig1}
\end{figure}

In an equilibrium system at low temperature, the collective excitation modes discussed in the previous section are thermally populated according to a Bose distribution with zero chemical potential. In the non-interacting limit, this Bogoliubov approach recovers the textbook prediction for the non-condensed density discussed in Secs. \ref{carusotto_sec2}. In the interacting case, in addition to these thermal fluctuations of the matter Bose field, a further contribution to the non-condensed fraction comes from the so-called quantum depletion of the condensate, i.e. quantum fluctuations due to virtual scattering of condensed particles into the non-condensed modes.

In spatial dimension $d<3$, the long-range order of the condensate is not stable against thermal fluctuations and is replaced by a so-called quasi-condensate, in agreement with the Hohenberg-Mermin-Wagner theorem of statistical physics~\cite{Huang}: while order is present up to intermediate length scales, upon increasing the distance it decays (at finite $T$) with an exponential (in $d=1$) or algebraic (in $d=2$) law. 
While this result was first discussed in the late 60's for equilibrium systems~\cite{Reatto,Popov}, a first mention in the non-equilibrium context was already present in the above-cited seminal work by Graham and Haken~\cite{Graham:ZPhys1970}: in this latter case, however, fluctuations did not have a thermal origin, but were a unavoidable consequence of the quantum nature of the field undergoing condensation.
Independently of this pioneering work, this result was rediscovered later on in Refs.~\cite{Wouters:PRB2006,Szymanska:PRL2006} and then extended to the critical region using renormalization-group techniques. Within this approach, several new features emerged: in $d=3$ novel critical exponents appear~\cite{3Dcritical} while the $d=2$ algebraic long-range decay of correlations is destroyed and replaced by a stretched exponential~\cite{2Dcritical}.

\section{How to quantitatively assess equilibrium?}

In most experiments on BEC in polariton and photon gases so far, a special effort was made in order to understand whether the system was thermalized or not. One can in fact expect that some effective thermalization should occur even in non-equilibrium regimes as soon as the thermalization time is shorter than the lifetime of the particles.
 In order to experimentally assess this {\em quasi-equilibrium} condition, the measured momentum and/or energy distribution of the non-condensed particle was typically compared to a Bose distribution.

The observation of such thermal distributions is to be expected in polariton and photon gases showing frequent collisions~\cite{klaers_schmitt_10,Kasprzak} as illustrated in the left panel of Fig.\ref{carusotto_fig1}. On the other hand, it was quite a surprise when the experiment~\cite{Bajoni} reported a momentum distribution with a thermal-like tail even in a lasing regime where the photons should not be thermalized ---  see the right panel of Fig.\ref{carusotto_fig1}. 
This observation cast some doubts on the interpretation of similar available experimental results; in particular, several authors have tried to develop alternative models to justify the observed thermal tail in terms of generalized, strongly non-equilibrium laser theories~\cite{fischer2012does,chiocchetta_carusotto_14}. 

This on-going debate calls for the identification of novel criteria to quantitatively assess the equilibrium vs. non-equilibrium nature of a system. A possible approach to this problem
will be the subject of the next sections.

\subsection{Fluctuation-dissipation theorems}
\label{sec:FDT}
Thermodynamical equilibrium is not only a property of the \emph{state} of a system, but also of its \emph{dynamics}. A remarkable consequence of equilibrium which involves dynamical quantities is the so-called fluctuation-dissipation theorem (FDT) \cite{kubo_66} which provides a relationship between the linear response of a system to an external perturbation of frequency $\omega$ and the thermal fluctuations of the same system at the same frequency $\omega$. 
While FDT relations hold for any pair of operators, in the following of this work we focus on the annihilation and creation operators $b_\kk$ and $b_\kk^\dagger$ of a $\kk\neq 0$  non-condensed mode of the intra-cavity photon/polariton field undergoing BEC in a spatially homogeneous geometry. To account for the particle number variation, it is convenient to include a chemical potential in the Hamiltonian, $H=H_0-\mu N$, so that frequencies $\omega$ are measured from the chemical potential. In the presence of a condensate, $\mu$ coincides with the oscillation frequency of the condensate mode, which in optical systems is observable as the condensate emission frequency $\omega_{BEC}$.

To state the FDT  theorem it is convenient to introduce the two functions
\begin{equation}
C_{bb^\dagger}(\kk,t-s) = \frac{1}{2} \left\langle \left\{ \bh_\kk(t) , \bhd_\kk(s) \right\}\right\rangle, \quad  \chi''_{bb^\dagger}(\kk,t-s)  = \frac{1}{2}\left\langle \left[\bh_\kk(t),\bhd_\kk(s)\right] \right\rangle,
\end{equation}
where the time dependence of the operator corresponds to their Heisenberg evolution under the system Hamiltonian $H$ and the average $\langle\dots \rangle$ is taken in a thermal equilibrium state at temperature $T$ with density matrix $\rho \propto \exp(-H/k_BT)$. In such a state, these correlations only depend on the time difference $t-s$ and we can define their Fourier transforms $C_{bb^\dagger}(\kk,\omega)$ and $\chi''_{bb^\dagger}(\kk,\omega)$. 
The explicit form of the FDT then reads:
\begin{equation}
C_{bb^\dagger}(\kk,\omega) = \coth\left(\frac{\omega}{2k_B T}\right) \chi''_{bb^\dagger}(\kk,\omega).
\eqname{carusotto_FD}
\end{equation}
In order to understand the physical content of this relation, we note that $\chi''_{bb^\dagger}(\kk,\omega)$ appearing in this relation is directly related to the imaginary part of the response function $\chi_{bb^\dagger}(\kk,\omega)$ of the system which quantifies the energy it absorbs from the weak perturbation~\cite{mahan_book_00}, i.e. $\chi''_{bb^\dagger}(\kk,\omega) = -\text{Im}[\chi_{bb^\dagger}(\kk,\omega)]$. As usual, $\chi_{bb^\dagger}(\kk,\omega)$ is defined as the Fourier transform of the linear response susceptibility $\chi_{bb^\dagger}(\kk,t)=-2i\theta(t)\,\chi''_{bb^\dagger}(\kk,t)$.

An alternative, fully equivalent formulation of the FDT \eq{carusotto_FD} is the so-called Kubo-Martin-Schwinger (KMS) condition \cite{kubo_57, martin_schwinger_59}, which in our example reads 
\begin{equation}
\label{eq:KMS}
S_{b^\dagger b}(\kk,-\omega) = \ee^{-\beta \omega} S_{bb^\dagger}(\kk,\omega),
\end{equation}
where ${S}_{bb^\dagger}(\kk,t) = \langle \bh_\kk(t) \bhd_\kk \rangle$ and ${S}_{b^\dagger b}(\kk,t) = \langle \bhd_\kk(t) \bh_\kk \rangle$.

The FDT has quite often been used to probe the effective thermalization of a system and to characterize the eventual departure from equilibrium \cite{cugliandolo_11,foini_cugliandolo_11,foini_cugliandolo_12}. In particular, given a pair of correlation functions, one can always define from \eq{KMS} an effective temperature $T_\text{eff}$ such that the functions satisfy a FDT: if the system is really at equilibrium, $T_\text{eff}$ has a constant value independently of $\kk$ and $\omega$ and equal to the thermodynamic temperature. On the other hand, if the system is out of equilibrium $T_\text{eff}$ will generically develop a non-trivial dependence on $\kk$ and $\omega$.

\subsection{Application to photon/polariton condensates}
\label{sec:photon_polariton}
Applying these ideas to the photon/polariton condensates discussed in the previous Sections provides a quantitative criterion to assess the equilibrium or non-equilibrium nature of the condensate: the protocol we propose consists in measuring different correlation and/or response functions and in checking if they satisfy the FDT. 

On the one hand, the correlation function $S_{b^\dagger b}$ can be related to the angle- and frequency-resolved photoluminescence intensity $\mathcal{S}(\kk,\omega)$ coming from the non-condensed particles via $\mathcal{S}(\kk,\omega+\omega_{BEC})=S_{b^\dagger b}(\kk,-\omega)$, where the condensate emission frequency $\omega_{BEC}$ plays the role of the chemical potential $\mu$ in the non-equilibrium context. On the other hand, $\chi''_{bb^\dagger}(\kk,\omega)$ is related to the imaginary part of the linear response to an external monochromatic field with momentum $\kk$ and frequency $\omega+\omega_{BEC}$. To be more specific, let us assume a two-sided cavity illuminated by an external classical field $E^\text{inc}_\kk(t)$ incident from the left. This field couples to the intra-cavity bosons through the Hamiltonian~\cite{RMP,walls_milburn_book_07}
\begin{equation}
H_\text{pump} = i \int \frac{\dd^d k}{(2\pi)^d}\, \left[ \eta^\text{l}_\kk E^\text{inc}_\kk(t) b_\kk^\dagger - \eta^{\text{l}*}_\kk E^{\text{inc}*}_\kk(t) b_\kk  \right],
\end{equation}
where $\eta^\text{l}_\kk$ ($\eta^\text{r}_\kk$) is the transmission amplitude of the left (right) mirror of the cavity. In the steady state the intra-cavity field vanishes $\langle b_\kk\rangle_\text{eq} =0$ and the incident classical field induces a perturbation 
%
$\langle b_\kk(\omega)\rangle =  i \eta^\text{l}_\kk\,\chi_{bb^\dagger}(\kk,\omega)\,E_\kk^\text{inc}(\omega)$,
%
where $b_\kk(\omega) = \int \dd t\, b_\kk(t) \ee^{i\omega t}$. From the boundary conditions at the two mirrors, the reflected and transmitted fields can be related to the intra-cavity field~\cite{RMP,walls_milburn_book_07} as:  
\begin{align}
E_\kk^\text{refl}(\omega) & =  \left[ 1 - i |\eta_\kk^\text{l}|^2 \,\chi_{bb^\dagger}(\kk,\omega) \right] E_\kk^\text{inc}(\omega), \label{eq:BC1}\\
E_\kk^\text{tr}(\omega) & = -i\eta_\kk^{\text{r}*}\,\eta^\text{l}_\kk\,\chi_{bb^\dagger}(\kk,\omega)\, E_\kk^\text{inc}(\omega).\label{eq:BC2}
\end{align}
Accordingly, from a measurement of the reflected or transmitted fields it is possible to reconstruct the response function $\chi''_{bb^\dagger}(k,\omega)$ through the formulas
\begin{equation}
\label{eq:chi_im_input_output}
\chi''_{bb^\dagger}(\kk,\omega) = \frac{1}{|\eta_\kk^\text{l}|^2}\left(1- \text{Re}\left[\frac{E^\text{refl}_\kk(\omega)}{E^\text{inc}_\kk(\omega)}\right] \right) = -\frac{1}{|\eta_\kk^\text{l}|^2} \textrm{Re}\left[\frac{\eta^{\text{l}*}_\kk}{\eta_\kk^{\text{r}*}}\frac{E^\text{tr}_\kk(\omega)}{E^\text{inc}_\kk(\omega)}\right],
\end{equation}
where both the amplitude and the phase of $E^\text{refl}_\kk$ and $E^\text{tr}_\kk$ can be measured with standard optical tools and the $\eta_\kk^\text{l,r}$ coefficients can be extracted from reflection and transmission measurements on the unloaded cavity.

At equilibrium, these quantities are related to the angle- and frequency-resolved luminescence spectrum $\mathcal{S}$ by the FDT
\begin{equation}
\chi''_{bb^\dagger}(\kk,\omega) =  \frac{1-\ee^{-\beta\omega}}{2}\,S_{bb^\dagger}(\kk,\omega)=\frac{e^{\beta\omega}-1}{2}\,\mathcal{S}(\omega_{BEC}+\omega,\kk):
\end{equation}
as both sides of this equation are experimentally measurable, any discrepancy is a signature of a non-equilibrium condition.

As a further verification, the interested reader may check that this FDT is satisfied for an empty cavity which is illuminated from both sides by thermal radiation at the same temperature.

\subsection{Application to some models of photon/polariton BEC}
As a final point, we will illustrate the behaviour of the FDT for two simple models of photon/polariton BEC. In doing this, one has to keep in mind that some of the approaches usually used to describe open quantum systems are intrinsically unable to correctly reproduce the FDT, so one must be careful not to mistake an equilibrium system for a non-equilibrium one just because of the approximations made in the theoretical model. In particular, every quantum master equation governed by a Lindblad super-operator always violates the FDT \cite{talkner_86, ford_o'connell_96}, even if the stationary solution has the form of a thermal density matrix: The reason of this pathology lies in the full Markovian approximation, which is inherent in the master equations \cite{gardiner_zoller_book_04}.

\subsubsection{Quantum Langevin model}

In Ref.~\cite{chiocchetta_carusotto_14} two of us proposed a simple model of non-equilibrium condensation based on a generalized laser model. The idea is to model the complex scattering processes responsible for condensation in terms of a spatially uniform distribution of population-inverted two-level atoms, which can emit light into the cavity mode. Once the population inversion is large enough, laser operation will occur into the cavity. The dynamics of quantum fluctuations on top of the coherent laser emission is then described by means of quantum Langevin equations for the non-condensed mode amplitudes.

\begin{figure}[htbp]
\centerline{\begin{tabular}{cc}
	\includegraphics[width=6cm]{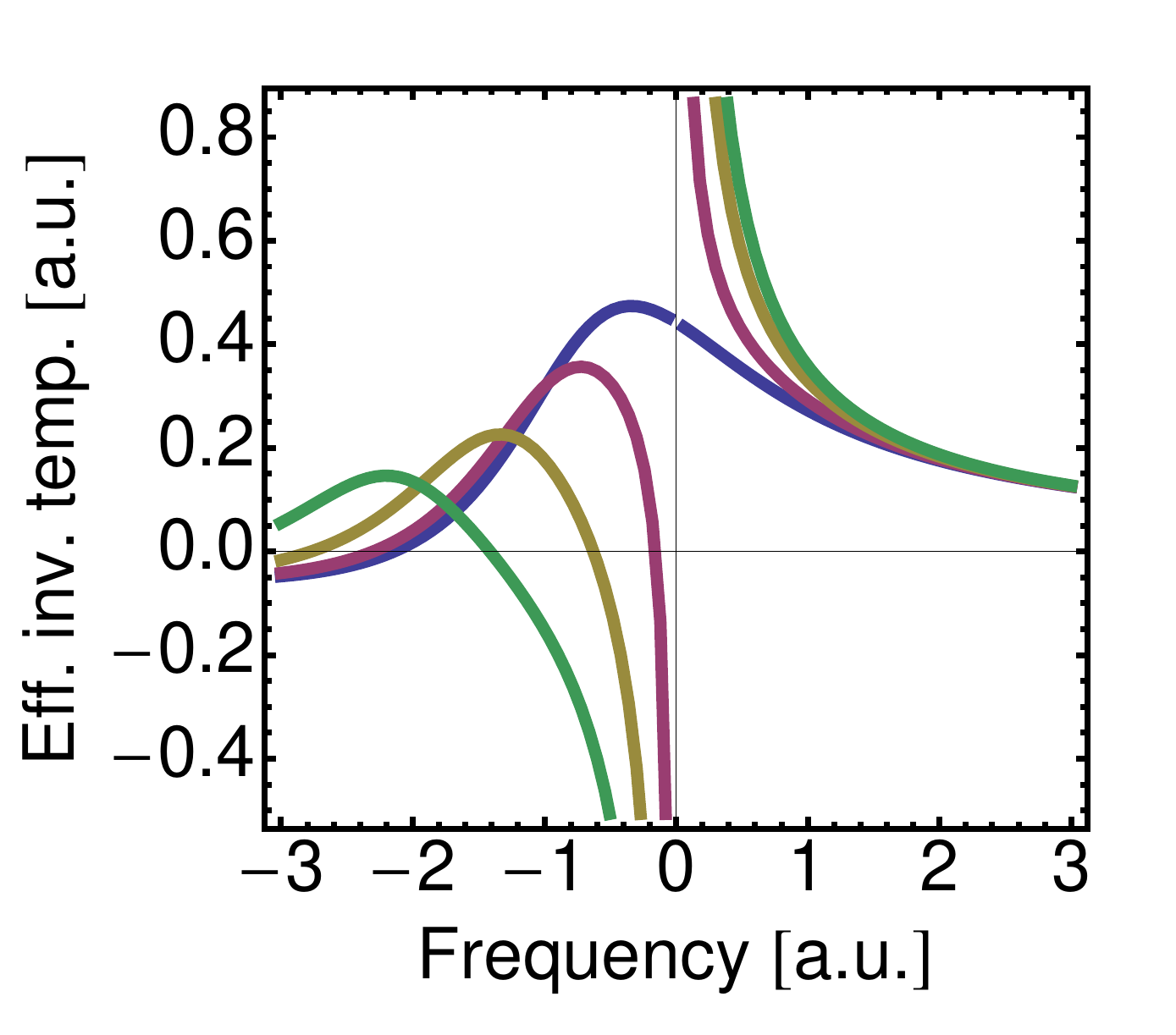} &
	\includegraphics[width=6cm]{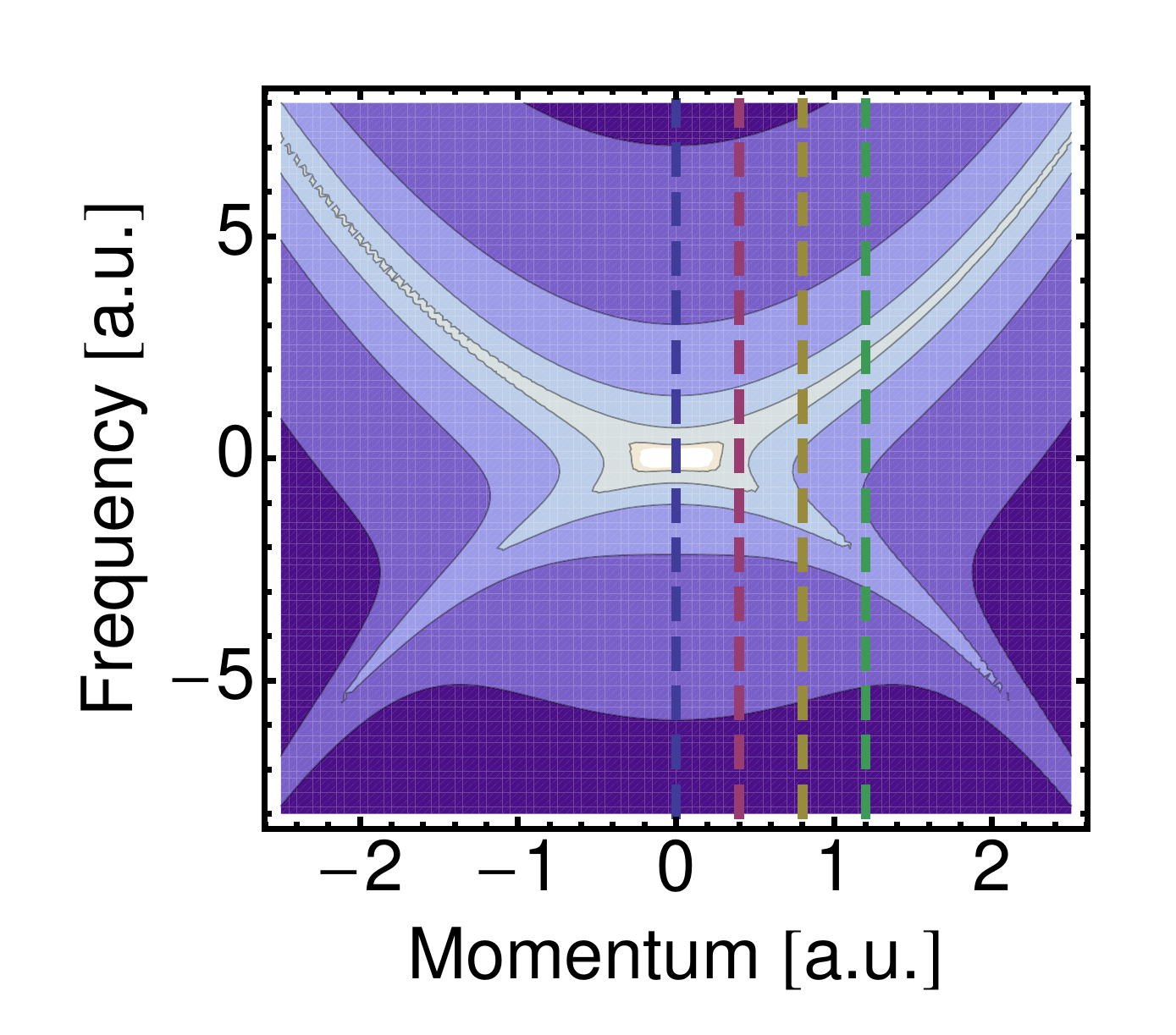}
\end{tabular}}
\caption{Left panel: Effective inverse temperatures vs frequency for the model of Ref.~\citep{chiocchetta_carusotto_14}. The various curves correspond to different values of the in-plane momenta $\kk$.
Right panel: Angle- and frequency-resolved photoluminescence intensity $\mathcal{S}(\kk,\omega_{BEC}+\omega)$ from Ref.~\citep{chiocchetta_carusotto_14}. The dashed vertical lines correspond to the in-plane momenta considered in the left panel.}
\label{carusotto_fig2}
\end{figure}

To check the effective lack of thermalization of the system, in Fig. \ref{carusotto_fig2} we study the $\omega$ and $\kk$ dependence of the effective inverse temperature $\beta_\text{eff}$, as extracted from the KMS relation \eqref{eq:KMS} using the quantum Langevin prediction for the correlation functions. The resulting $\beta_\text{eff}(\kk,\omega)$ strongly depends on both $\omega$ and $\kk$ and becomes even negative in some regions: all these features are a clear signature of a very non-equilibrium condition.         
\subsubsection{Non-Markovian toy model}

The situation is much more intriguing for the model recently proposed in~\cite{kirton_keeling_13,kirton_keeling_14} to study the photon BEC experiments of Ref.\cite{klaers_schmitt_10}: clear signatures of a thermal distribution of the non-condensed cloud were observed as soon as the thermalization rate under the effect of repeated absorption and emission cycles by the dye molecules becomes comparable to the photon loss rate. On the other hand, when thermalization is too slow, the thermal-like features break down and the system reproduces the non-equilibrium physics of a laser.

\begin{figure}[htbp]
\centerline{\begin{tabular}{cc}
	\includegraphics[width=6cm]{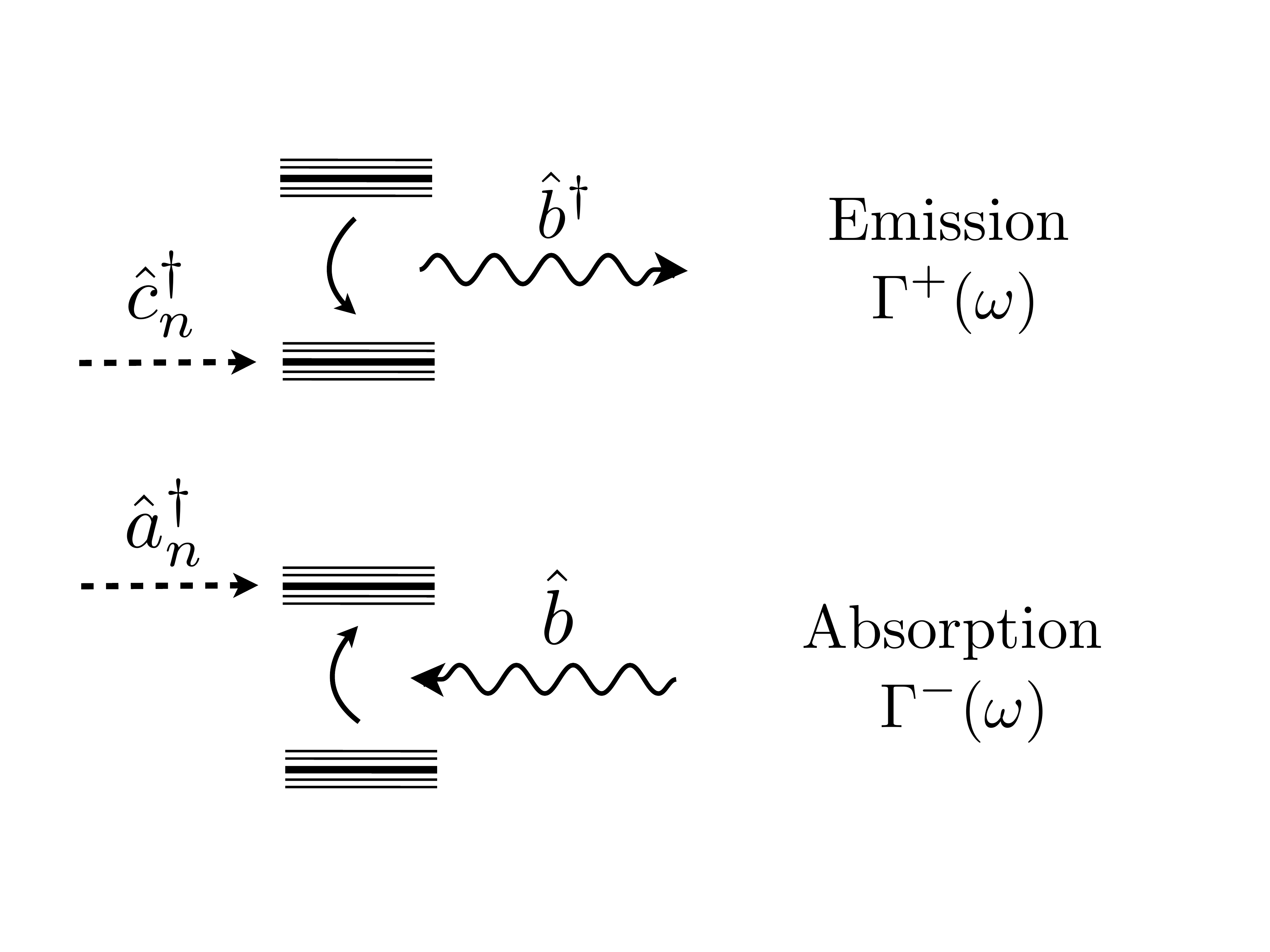} &
	\includegraphics[width=5cm]{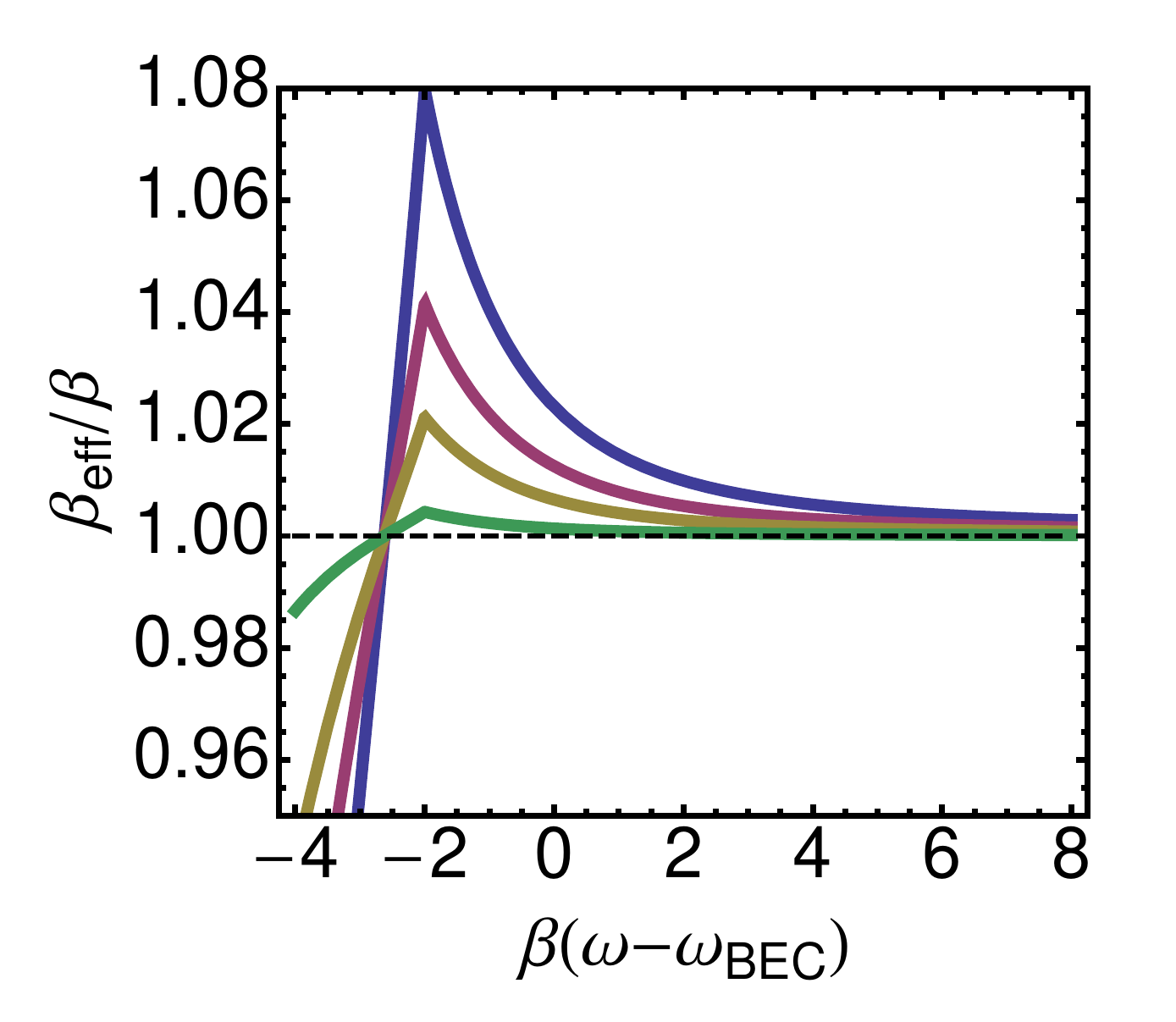}
\end{tabular}}
\caption{Left panel: Sketch of the energy levels under consideration. Right panel: Effective inverse temperatures $\beta_\text{eff}$ for $\beta(\omega_{BEC}-\bar{\omega}) = 2$ and for different values of $\kappa/\gamma^-=$ $0.2$ (blue), $0.1$ (red), $0.05$ (yellow), $0.01$ (green). }
\label{carusotto_fig3}
\end{figure}

In order to investigate how this crossover affects the FDT, we introduce a non-Markovian toy model which extends the theory in~\cite{kirton_keeling_13,kirton_keeling_14} to avoid spurious effects due to the Markov approximation.
For simplicity, we consider a single non-condensed mode of frequency $\omega_c$ described by operators $\bh, \bhd$. The frequency-dependent absorption and amplification by the dye molecules is modeled in terms of two distinct baths of harmonic oscillators $\ah_n, \ahd_n$ and $\ch_n, \chd_n$ with frequencies $\omega_n^-$ and $\omega_n^+$, respectively:
\begin{equation}
H  = \omega_c \bhd \bh + \sum_{n} (\omega_n^- \ahd_n \ah_n + \omega_n^+ \chd_n \ch_n) + \sum_n \left(\eta^-_n \bh \ahd_n + \eta^+_n \bhd \chd_n + \text{H.c.}\right) 
\eqname{HKK}.
 \end{equation}
The baths take into account the two processes pictorially represented in the left panel of Fig.~\ref{carusotto_fig3}, in which the photon absorption and emission processes are associated to the creation/destruction of ro-vibrational phonons. An analogous absorbing bath is used to model cavity losses due to the imperfect mirrors.    

According to the usual quantum-Langevin theory \citep{gardiner_zoller_book_04}, we solve the Heisenberg equations of motions for $\ah_n, \ahd_n$ and $\ch_n, \chd_n$ and replace the formal solution into the Heisenberg equation for $\bh$, which takes the simple form
\begin{equation}
\label{eq:QL-1}
\frac{\dd \bh}{\dd t} = -\left(i\omega_c +\frac{\kappa}{2}\right)\bh - \int_{-\infty}^{+\infty} \dd t'\, [\Gamma^-(t-t')-\Gamma^+(t-t')]\bh(t') +F,
\end{equation}
where $\kappa$ is the decay rate of the cavity photon, $F= F^\kappa + F^- + F^+$ is the total noise operator and the memory kernels $\Gamma^\pm$ are defined as $\Gamma^\pm(t) = \theta(t) \int_{-\infty}^{+\infty} \dd \omega\, \rho^\pm(\omega) \ee^{\pm i\omega t}/(2\pi) $, 
in terms of the absorption and emission spectral functions $\rho^\pm(\omega) = 2\pi \sum_n |\eta^\pm_n|^2\delta(\omega - \omega_n^\pm) $.

Given the form of the bath-system coupling \eq{HKK} and of the memory kernels $\Gamma^\pm(t)$, absorption (viz. amplification) of a photon at $\omega_c$ is proportional to $\rho^-(\omega)$ (viz. $\rho^+(-\omega)$). The Kennard-Stepanov (KS) relation between the absorption and emission spectra from molecules in thermal contact with an environment at inverse temperature $\beta$ then translates into 
\begin{equation}
\label{eq:Kennard-Stepanov}
\rho^+(-\omega) = \mathcal{C}\, \ee^{-\beta \omega} \rho^-(\omega),
\end{equation}
the constant $\mathcal{C}$ depending on the pumping conditions, e.g. the fraction of molecules in the ground and excited electronic states. 
In what follows, we assume all baths to be initially in their vacuum state, so as to model irreversible absorption and emission processes. In this regime, the structure factors read:
\begin{equation}
S_{b^\dagger b}(-\omega) =  \frac{ \rho^+(-\omega) }{[\omega - \Sigma(\omega)]^2 + \Gamma^2_T(\omega)} , \quad S_{bb^\dagger}(\omega) =  \frac{\kappa + \rho^-(\omega)}{[\omega  - \Sigma(\omega)]^2 + \Gamma^2_T(\omega)},
\eqname{Ss} \end{equation}
where $\Sigma(\omega) = \omega_c - \text{Im}[\Gamma^-(\omega)] + \text{Im}[\Gamma^+(\omega)]$ takes into account the Lamb-shift induced by the baths and the total relaxation rate is $\Gamma_T(\omega) = \kappa/2 + \textrm{Re}[\Gamma^-(\omega)] - \textrm{Re}[\Gamma^+(\omega)]>0$. 
For concreteness, we choose the forms
\begin{eqnarray}
 \rho^-(\omega) &=& \gamma^- \{ [n(\omega-\bar{\omega})+1]\,\theta(\omega-\bar{\omega})+ n(\bar{\omega}-\omega)\,\theta(\bar{\omega}-\omega)\}, \label{eq:rate-} \\
 \rho^+(-\omega) &=& \gamma^+ \{ n(\omega-\bar{\omega})\,\theta(\omega-\bar{\omega})+ [n(\bar{\omega}-\omega)+1]\,\theta(\bar{\omega}-\omega)\}, \label{eq:rate+}  
 \end{eqnarray}
 modelling phonon-assisted absorption on a molecular line at $\bar{\omega}$, with $\gamma^\pm$ proportional to the molecular population in the ground and excited states and $n(\omega) = [\exp(\beta \omega) - 1]^{-1}$ . It is straightforward to check that these forms for $\rho^\pm$ indeed satisfy the KS relation \eq{Kennard-Stepanov} with $\mathcal{C}=e^{\beta\bar{\omega}} \gamma^+/\gamma^-$. Dynamical stability of the condensate imposes the further condition $\kappa + \rho^-(\omega_{BEC})=\rho^+(-\omega_{BEC})$, which translates into $\mathcal{C}=e^{\beta\omega_{BEC}}[1+ \kappa/\rho^-(\omega_{BEC})]$.
 

As discussed in Sec.\ref{sec:FDT}, the frequencies appearing in the KMS condition \eqref{eq:KMS} are measured from the chemical potential. Even with this rescaling, it is immediate to see that the structure factors do not generally satisfy the KMS condition \eqref{eq:KMS} signalling a non-equilibrium behaviour. 
However, in the plots shown in the right panel of Fig. \ref{carusotto_fig3} of the KMS effective inverse temperature 
\begin{equation}
\beta_{\rm eff}(\omega) = \frac{\log\left(\frac{\kappa+\rho^-(\omega)}{\rho^+(-\omega)}\right)}{\omega-\omega_{BEC}} =
\beta +\frac{1}{\omega-\omega_{BEC}} \log \left[\frac{ 1 + \kappa/\rho^-(\omega) }{ 1 + \kappa/\rho^-(\omega_{BEC}) }\right]
\end{equation}
one easily sees that an effective equilibrium at $\beta$ is recovered in the $\kappa\to 0$ limit where the KS condition \eq{Kennard-Stepanov} makes the KMS condition to be trivially fulfilled. Physically, if the repeated absorption and emission cycles by the molecules are much faster than cavity losses, the KS condition imposes a full thermal equilibrium condition in the photon gas. 

\section{Conclusions}
After reviewing the most intriguing novel features of non-equilibrium BEC and laser operation as compared to textbook BEC, we have proposed and characterized a quantitative criterion to experimentally assess the equilibrium vs. non-equi\-li\-brium nature of a condensate. 
This criterion has been applied to a strongly non-equilibrium model of condensation inspired to the semi-classical theory of laser and to a simple non-Markovian model of the photon BEC: provided photons undergo repeated absorption-emission cycles before being lost, the photon gas can inherit the thermal condition of the dye molecules.
With respect to static properties, such as the momentum distribution, so far considered in experiments, our criterion based on fluctuation-dissipation relations imposes stringent conditions also on the dynamical properties of the gas: its experimental implementation appears feasible with state-of-the-art technology and would give a conclusive evidence of thermal equilibrium in the gas.


\bibliography{Carusotto_CUP_References}
\bibliographystyle{cambridgeauthordate}

\end{document}